\documentclass[reprint,superscriptaddress,amsmath,amssymb,floatfix,aps,pra]{revtex4-2}
\usepackage{stix2}
\usepackage{graphicx}
\usepackage{xcolor}
\usepackage{bm}
\usepackage{hyperref}
\usepackage{braket}
\usepackage{ulem}
\hypersetup{colorlinks=true,linkcolor=blue,citecolor=blue,urlcolor=magenta}
    \newcommand{\bi}[1]{\bm{#1}}        
    \newcommand{\pd}{\partial}      
    \newcommand{\dd}{\mathrm{d}}    
    \newcommand{\ii}{\mathrm{i}}    
    \newcommand{\ee}{\mathrm{e}}    
    \newcommand{\abs}[1]{\lvert #1 \rvert}  
\begin{document}

\title{\textsc{Ionization Induced by the Ponderomotive Force \\ in Intense and High-Frequency Laser Fields}}
\author{Mingyu Zhu}
    \affiliation{State Key Laboratory of Precision Spectroscopy, East China Normal University, Shanghai 200241, China}
    \affiliation{School of Physics and Electronic Science, East China Normal University, Shanghai 200241, China}
\author{Yuxiang Liu}
    \affiliation{School of Physics and Electronic Science, East China Normal University, Shanghai 200241, China}
\author{Chunli Wei}
    \affiliation{School of Physics and Electronic Science, East China Normal University, Shanghai 200241, China}
\author{Hongcheng Ni}
    \email{hcni@lps.ecnu.edu.cn}
    \affiliation{State Key Laboratory of Precision Spectroscopy, East China Normal University, Shanghai 200241, China}
    \affiliation{Collaborative Innovation Center of Extreme Optics, Shanxi University, Taiyuan, Shanxi 030006, China}
    \affiliation{NYU-ECNU Joint Institute of Physics, New York University at Shanghai, Shanghai 200062, China}
\author{Qi Wei}
    \email{qwei@admin.ecnu.edu.cn}
    \affiliation{State Key Laboratory of Precision Spectroscopy, East China Normal University, Shanghai 200241, China}
\date{April 19, 2023}
\begin{abstract}
    Atomic stabilization is a universal phenomenon that occurs when atoms interact with intense and high-frequency laser fields.
    In this work, we systematically study the influence of the ponderomotive (PM) force, present around the laser focus, on atomic stabilization.
    We show that the PM force could induce tunneling and even over-barrier ionization to the otherwise stabilized atoms.
    Such effect may overweight the typical multiphoton ionization under moderate laser intensities.
    Our work highlights the importance of an improved treatment of atomic stabilization that includes the influence of the PM force.
\end{abstract}
\maketitle

\section{Introduction}
\label{sec:intro}

The interaction of atoms with intense and high-frequency laser fields has given rise to a variety of topics and extensive investigations for decades \cite{gavrila_atomic_2002, eberly_atomic_1993, gavrila_atoms_1992}.
The atomic stabilization against ionization is one of the most interesting phenomena,
in which the ionization probability decreases with increasing field intensity.
Perfect interpretation has been made on this phenomenon
by switching from the laboratory frame to an accelerated frame called the Kramers-Henneberger (KH) frame \cite{henneberger_perturbation_1968},
where the nucleus quivers under the laser field and forms a time-dependent potential.
The high-frequency Floquet theory (HFFT) \cite{gavrila_free-free_1984} is the main theoretical framework to be applied on this topic.
The lowest order of HFFT yields the prediction of adiabatic stabilization in the high-frequency regime,
in which the time-dependent potential the electron experiences becomes static by time-averaging the dynamic one,
and the distorted potential was named ``dressed'' potential, depending only on the quiver amplitude $\alpha_0$ of the electron
    \cite{gersten_shift_1976, gavrila_free-free_1984, pont_dichotomy_1988, pont_atomic_1989}.
Atoms in states of such potential are called KH atoms, forming a relatively stable structure \cite{pont_atomic_1988, pont_stabilization_1990}.
In linearly polarized laser field,
the dressed potential can be viewed as a potential formed by a charged line segment twice the length of $\alpha_0$ along the polarized direction,
where the charge density peaks at the two end points of the segment.
For large $\alpha_0$, the ground-state wave function under such a potential splits into two non-overlapping parts,
exhibiting a ``dichotomy'' structure \cite{pont_dichotomy_1988, pont_radiative_1990}, which even leads to double-slit interference like in a diatomic molecule \cite{he_youngs_2020}.
Several theoretical and experimental studies have been performed in attempting to confirm the existence of KH atoms
    \cite{morales_imaging_2011, wei_pursuit_2017, matthews_amplification_2018, he_robust_2021}.
\par

Study on atomic stabilization has attracted considerable attention since its discovery.
The past theoretical studies mainly focused on two forms of stabilization,
namely adiabatic stabilization and dynamic stabilization.
The former is related to multiphoton ionization initially derived from HFFT, where the ionization rate reduces with increasing laser intensity and frequency
    \cite{pont_stabilization_1990, pont_multiphoton_1991_i, pont_multiphoton_1991_ii, jiang_signatures_2022}.
The latter is usually studied by numerically solving the time-dependent Schrödinger equation (TDSE) that takes account of full laser-atom interaction,
thereby establishing the relation between the survival probability of atoms and laser parameters
    \cite{su_dynamics_1990, kulander_dynamic_1991, pont_numerical_1991, pont_observability_1991, reed_suppression_1991, grobe_packet_1992, vos_effective_1992, huens_dynamical_1993, gajda_ionization_1994, patel_stabilization_1998, liang_intensity-dependent_2021}.
As for experimental studies, several experiments have also confirmed the existence of stabilization
    \cite{de_boer_indications_1993, de_boer_adiabatic_1994, van_druten_adiabatic_1997, toyota_slow_2009}.
However, certain factors may lead to breakdown of atomic stabilization.
For example, non-adiabatic effects brought by turn-on and turn-off of an ultra-short laser pulse may induce ``shake-off'' ionization
    \cite{grobe_packet_1992, vivirito_adiabatic_1995, patel_effect_1999},
non-dipole and relativistic effects at high intensities may also contribute to destabilization
    \cite{keitel_monte_1995, vazquez_de_aldana_magnetic-field_1999, kylstra_breakdown_2000, vazquez_de_aldana_atoms_2001, forre_exact_2005, forre_nondipole_2006, simonsen_magnetic-field-induced_2015, lindblom_semirelativistic_2018}.
By far, the effect of ponderomotive (PM) force on atomic stabilization has not been studied.
For a plane-wave field, the PM energy is spatially constant and can be omitted as it should be.
For a typical laser pulse, however, the intensity follows a Gaussian profile at the focus,
where electrons will experience PM forces from the non-uniform spatial distribution of intensity
    \cite{eichmann_acceleration_2009, wang_neutral_2016, wei_pursuit_2017, wei_dynamics_2023}.
A recent study reveals that even a slight component of the PM force along the polarization direction can break the symmetry and change the electronic structure of KH atoms \cite{zhang_symmetry_2020}.
In this article, we investigate the effect of the PM force on atomic stabilization of KH atoms.
We show that the PM force could induce tunneling and even over-barrier ionization to the otherwise stabilized atoms, thereby leading to breakdown of stabilization.

\par

This article is organized as follows.
In Sec.~\ref{sec:PMI} we detail the ionization rate induced by the PM force, and discuss its dependence on system parameters.
In Sec.~\ref{sec:MPI} we review the multiphoton ionization of KH atoms and show how the PM force influences its characteristics.
In Sec.~\ref{sec:cp_PMI_MPI} we present a comparison of PM-force-induced ionization and multiphoton ionization within a typical laser setting.
Conclusions are given in Sec.~\ref{sec:concl}.
Atomic units are used throughout unless stated otherwise.

\section{Ponderomotive-Force-Induced Tunneling Ionization and Over-Barrier Ionization}
\label{sec:PMI}
\subsection{Ponderomotive Force and its Effects on KH Atoms}
\label{subsec:PMForce}

In the simplest case we consider atoms with single active electron (SAE).
Under the SAE approximation, the TDSE in Lab frame is written as
\begin{equation}
    H_\text{L}\varPsi_\text{L} = \left\{ \left( \bi{p}+\frac{1}{c}\bi{A}(\bi{r},t)\right)^{2}+V(\bi{r}) \right\}\varPsi_\text{L} = \ii\pd_t\varPsi_\text{L},
\end{equation}
where $\bi{p}$ is the momentum,
$\bi{A}(\bi{r},t)$ is the vector potential of the laser field,
and $V(\bi{r})$ is the Coulomb potential of the nucleus.
Applying the KH transformation
\begin{equation}
    U_\text{KH} = \exp\left[ \ii\bi{\alpha}(\bi{r},t) \cdot\bi{p} \right],
\end{equation}
we transform the reference frame from the laboratory frame to an accelerated frame called the KH frame,
where
\begin{equation}
    \bi{\alpha}(\bi{r},t) = \frac{1}{c} \int^t \bi{A}(\bi{r},\tau)\dd\tau
\end{equation}
describes the quiver motion of a classical electron in the laser field.
The transformed wave function reads
\begin{equation}
    \varPsi_\text{KH}(\bi{r},t) = U_\text{KH}\varPsi_\text{L}(\bi{r},t) = \varPsi_\text{L}(\bi{r}-\bi{\alpha}(\bi{r},t) ,t).
\end{equation}
The Hamiltonian, after neglecting higher order terms \cite{forre_exact_2005}, is rewritten as
\begin{equation}
    H_\text{KH} = \frac{\bi{p}^2}{2} + V(\bi{r}+\bi{\alpha}(\bi{r},t) ) + \frac{\bi{A}^{2}(\bi{r},t)}{2c^{2}},
    \label{eq:KH_Hamiltonian}
\end{equation}
where the first two terms defines the KH atom we are familiar with.
\par

Now we consider the last term of Eq.~(\ref{eq:KH_Hamiltonian}), namely $\bi{A}^2(\bi{r},t)/2c^2$.
For a laser pulse linearly polarized in $x$ direction and propagating along $z$ direction,
the vector potential is expressed as
 \begin{equation}
 \bi{A}\left(\bi{r},t\right)=\frac{c}{\omega}\mathcal{E}_0(\bi{r},t)\sin\left(\omega t-kz\right)\hat{\bi{x}}.
 \end{equation}
with $\mathcal{E}_0(\bi{r},t)$ the electric field amplitude, $\omega$ the laser frequency, $k$ the wave vector, and $\hat{\bi{x}}$ the polarization direction. Taking the cycle-average of $\bi{A}^2(\bi{r},t)/2c^2$ gives rise to the PM potential
\begin{equation}
    H_\text{PM} = \left<\frac{\bi{A}^{2}(\bi{r},t)}{2c^{2}}\right> = \abs{\mathcal{E}_0(\bi{r},t)}^2 / 4\omega^2.
\end{equation}
For a plane wave, $H_\text{PM}$ is constant everywhere, and therefore can be ignored.
For a typical focused laser beam, however, the the non-uniform spatial distribution of the PM potential at the laser focus result in the PM force
\begin{equation}
    \bi{F}_\text{PM} = -\bi{\nabla} H_\text{PM} = -\frac{1}{4\omega^2} \bi{\nabla}\abs{\mathcal{E}_0(\bi{r},t)}^2.
\end{equation}
To present an example,
we assume a linearly polarized laser beam with Gaussian spatial
intensity distribution,
propagating along the $z$ axis, which can be specified in cylindrical coordinates,
\begin{equation}
    I(\bi{r},t) = \abs{\mathcal{E}_0(\bi{r},t)}^2 = I_0 \left( \frac{w_0}{w(z)} \right)^2 \exp\left( -\frac{2r^2}{w(z)^2} \right)f(\eta),
    \label{eq:laser_profile}
\end{equation}
where $I_0$ is the peak intensity and $w(z)=w_0\sqrt{1+(z/z_\text{R})^2}$ with $w_0$ the beam waist and $z_\text{R}=\pi w_0^2/\lambda$ the Rayleigh length.  $f\left(\eta\right)=\exp[-\left(\eta/c\tau\right)^{2}]$ with $\eta=ct-z$, is a Gaussian laser pulse envelope profile of optical intensity with a finite pulse duration $\tau =\tau _\text{FWHM}/2\sqrt{\ln2}$, where $\tau_\text{FWHM}$ is the full width at half maximum of the laser pulse. 
For typical lasers, $z_\mathrm{R}$ is more than 100 times larger than $w_0$,
which implies that the gradient of the laser field and the corresponding PM force in the propagation direction are much smaller than that in the radial direction, and thus negligible \cite{eichmann_acceleration_2009}.
\par

In this article we take the hydrogen atom as the simplest example.
According to HFFT, in the high-frequency limit, the system can be viewed as quasi-stationary,
and dynamics of the electron can be illustrated by the stationary Schrödinger equation
\begin{equation}
    \left[ \frac{\bi{p}^2}{2} + V_0(\alpha_0;\bi{r}) - \bi{F}_\text{PM}\cdot\bi{r} \right]\varPsi = E\varPsi,
    \label{eq:TISE}
\end{equation}
where $V_0(\alpha_0;\bi{r})$ is the dressed potential,
which is defined as the time average of the oscillating potential $V(\bi{r}+\bi{\alpha}(\bi{r},t))$:
\begin{equation}
    V_0(\alpha_0;\bi{r}) = \Braket{V(\bi{r}+\bi{\alpha}(\bi{r},t))} = -\frac{1}{2\pi} \int_0^{2\pi/\omega} \frac{\dd t}{\abs{\bi{r}+\bi{\alpha}(\bi{r},t)}}.
\end{equation}
Designating $\bi{\alpha}(\bi{r},t)=\alpha_0\cos{\omega t}\hat{\bi{x}}$, with the quiver amplitude $\alpha_0=\abs{\mathcal{E}_0(\bi{r},t)}/\omega^2$,
the potential is expressed as
\begin{equation}
    V_0(\alpha_0;\bi{r}) = -\frac{2}{\pi} \frac{1}{\sqrt{r_+ r_-}} \mathrm{K}\left( \sqrt{(1-\hat{\bi{r}}_+ \cdot \hat{\bi{r}}_-)/2} \right),
\end{equation}
where $\bi{r}_\pm = \bi{r}\mp\alpha_0\hat{\bi{x}}$ are the position vectors relative to the two end points of the quiver motion, and $\mathrm{K}(k)$ is the complete elliptic integral of the first kind.
Since we are interested in situations where $\alpha_0\sim 10^2$ a.u.,
which is a quiver amplitude large enough that even for visible light, the high frequency requirement $\omega\gg I_\text{p}$ ($I_\text{p}$ is the ionization potential) is well satisfied.
\par

\begin{figure*}
    \includegraphics[width=15cm]{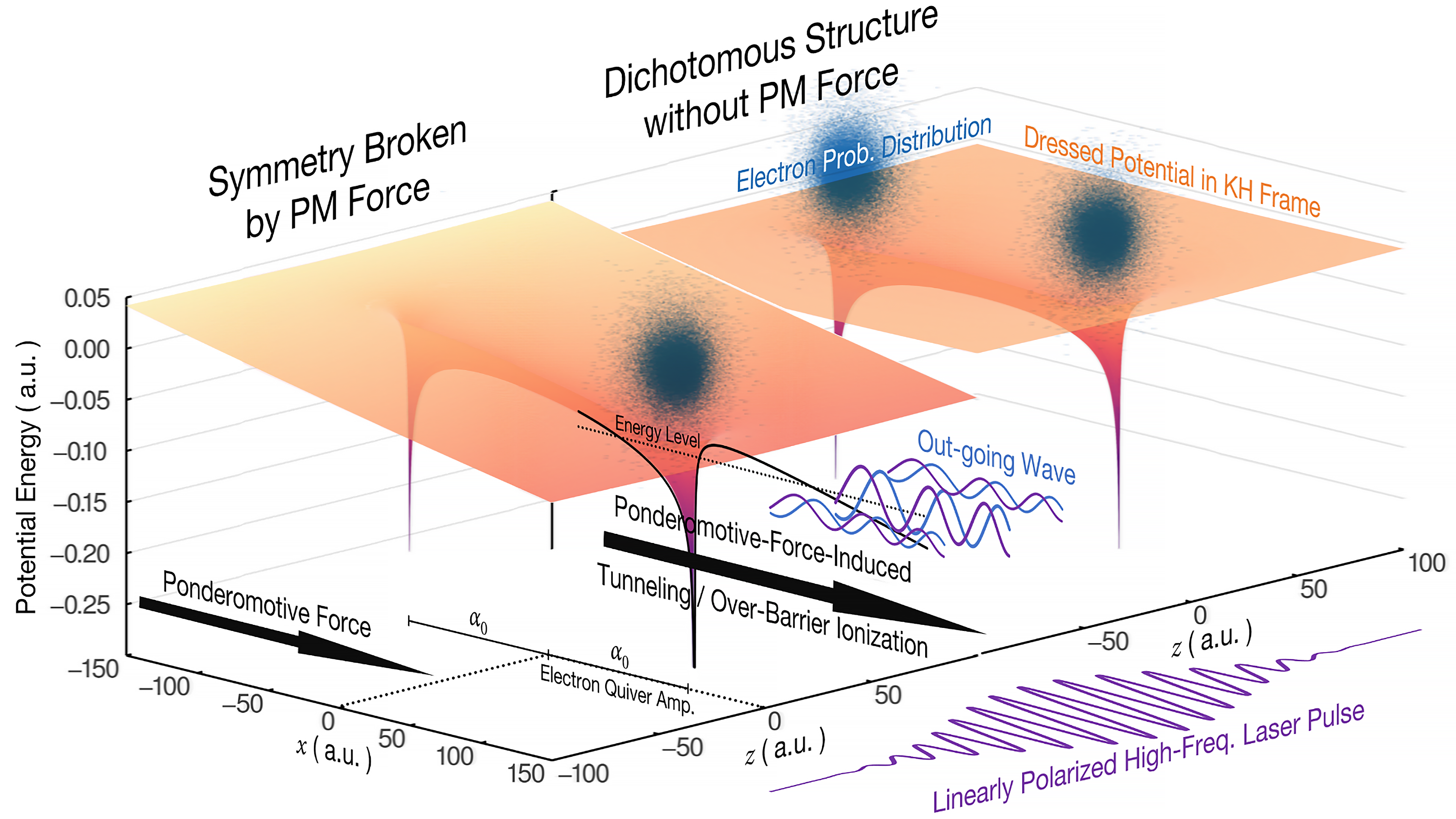}
    \caption{
        Effect of the ponderomotive (PM) force on the structure of the KH atom in a linearly polarized laser field.
        PM force may break the symmetric structure of the KH atom,
        and further lead to tunneling ionization or even over-barrier ionization.
        }
    \label{fig:KH_PM_Effect}
\end{figure*}

Despite the fact that PM forces are usually much weaker compared to that from the laser field,
they could have substantial influence on KH atoms' structure.
The component of the PM force along the polarization direction will break the symmetric dichotomous structure of KH atom,
where the electron tends to stay near the ``downstream'' end point, exhibiting a single-lobe structure.
Under the circumstances where the PM force is relatively strong, it would bend the KH potential,
forming a potential barrier, which the electron may go across and escape to the continuum,
resulting in tunneling ionization or over-barrier ionization.
Fig.~\ref{fig:KH_PM_Effect} presents a schematic sketch of the effects of the PM force on the KH atom.
\par

\subsection{Ionization Rate}
\label{subsec:IonRate}

In order to obtain the stationary ionization rate of KH atoms induced by the PM force,
we implement a numerical solution of the TDSE.
\par

In our scheme, the system evolves under the TDSE
\begin{equation}
    \ii\pd_t\varPsi = \left[ -\frac12\nabla^2 + V_0(\alpha_0;\bi{r}) - \Lambda(t)\bi{F}_\text{PM}\cdot\bi{r} \right]\varPsi,
    \label{eq:TDSE}
\end{equation}
where a smooth turn on
\begin{equation}
    \Lambda(t)=
    \begin{cases}
        0,                  &t<0, \\
        \sin^2 \pi t/2T,    &0\leqslant t\leqslant T, \\
        1,                  &t>T, \\
    \end{cases}
\end{equation}
is implemented to ensure the system go through adiabatic transition from the field-free ($F_\text{PM}=0$) initial state to the final state of stable ionization under the designated PM force.
The initial state is the ground eigenstate (1s state) of the field-free KH Hamiltonian, obtained through imaginary-time evolution.
It's noteworthy that the purpose of our scheme is to obtain the stationary ionization rate induced by the PM force,
and its dependence on $\alpha_0$, $F_\mathrm{PM}$ and $\theta$.
Thus Eq.~(\ref{eq:TDSE}) is not used to mimic the exact physical dynamics of the KH atoms,
but to calculate the stationary ionization rate induced by the PM force.
The TDSE simulation is carried out on a three-dimensional Cartesian grid,
calculation of the initial state and propagation of the wave function are carried out using the split-operator Fourier method \cite{feit_solution_1982}.
To avoid unphysical reflection on the boundaries, absorbing boundaries of the $\cos^{1/6}$ shape is applied after each time step propagation to suppress the out-going wave approaching the grid border.
The ionization rate is defined as the surface integral of the probability current density
\begin{equation}
    \Gamma(t) = \int_\Sigma \bi{J}\cdot\dd\bi{S} = \int_\Sigma \frac{\ii}{2}\left( \varPsi\nabla\varPsi^* - \varPsi^*\nabla\varPsi \right)\cdot\dd\bi{S}.
\end{equation}
Soon after $\Lambda(t)$ reaches 1 and the system stabilizes, we have $\Gamma(t)\propto\exp\left(-\Gamma_0 t\right)$,
thus the stationary ionization rate can be derived from
\begin{equation}
    \Gamma_0 = -\frac{\dd\ln{\Gamma(t)}}{\dd t}.
\end{equation}
\par

\begin{figure*}
    \includegraphics[width=17.8cm]{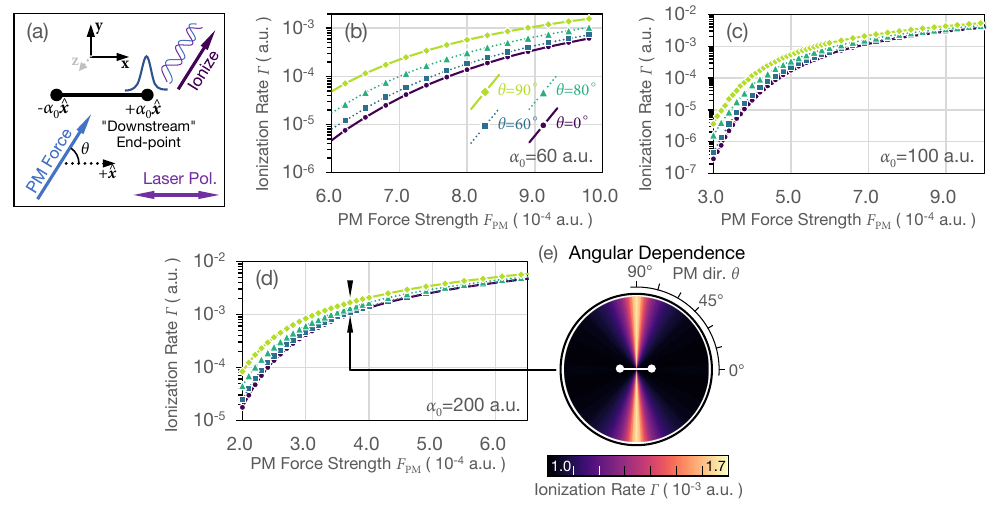}
    \caption{
        PM-force-induced ionization rates obtained by numerically solving the TDSE.
        (a) The $x$ axis denotes the laser's polarization direction,
        and $\theta$ ($0^\circ\leqslant\theta\leqslant 90^\circ$) represents the angle between the PM force and the $+x$ direction.
        For $\theta\neq 90^\circ$, the end point $+\alpha_0 \hat{\bi{x}}$ is the ``downstream'' one,
        i.e., $V_0(+\alpha_0 \hat{\bi{x}})<V_0(-\alpha_0 \hat{\bi{x}})$.
        In most situations, the electron wave function concentrates near the ``downstream'' end point.
        The dependence of the ionization rate on $F_\text{PM}$ and $\theta$ are presented in
        panel (b) with $\alpha_0=60\ \text{a.u.}$, panel (c) with $\alpha_0=100\ \text{a.u.}$, and panel (d) with $\alpha_0=200\ \text{a.u.}$
        The angular dependence of the ionization rate under the parameter $\alpha_0=200\ \text{a.u.}$ and $F_\text{PM}=3.7\times 10^{-4}\ \text{a.u.}$ is presented in panel (e).
        }
    \label{fig:IonRate}
\end{figure*}

Fig.~\ref{fig:IonRate} presents the static ionization rates obtained by numerically solving the TDSE [Eq.~(\ref{eq:TDSE})] following the above procedure, where Fig.~\ref{fig:IonRate}(a) is a sketch of the geometry of the problem.
\par

As displayed in Fig.~\ref{fig:IonRate}(b-d), the $F_\text{PM}$-dependence of the ionization rate is obvious:
For fixed $\alpha_0$, the ionization rate increases exponentially as $F_\text{PM}$ increases since the potential barrier gets thinner and lower for larger $F_\text{PM}$.
On the other hand, for a certain laser frequency,
the PM-free ground-state ionization potential $I_\text{p}=\abs{E_0}$ decreases and the PM force increases dramatically with increasing $\alpha_0$,
the PM-force-induced ionization is therefore much more common for larger $\alpha_0$.
Since $\alpha_0\propto\sqrt{I}$ and $F_\text{PM}\propto I$ ($I$ is the laser intensity),
ionization induced by the PM force would rapidly grow as the laser intensity increases,
eventually turning into a dominant factor that determines the lifetime of the KH atom.
\par

From Fig.~\ref{fig:IonRate}(e) we could further get a glimpse of the dependence of the ionization rate on the direction of the PM force:
The ionization rate peaks when the PM force is perpendicular to the laser polarization direction,
and decreases as the PM force rotates towards the polarization direction.
This trend can be understood when we look closely into the potential near the end points.
As will be shown in Eq.~(\ref{eq:EndPtVicinityApprox2}) in Sec.~\ref{subsec:ScaledIonRate}, the angular dependence of the potential near the end points follows as
\begin{equation}
    \tilde{V}_0 \propto -\mathrm{K}\left( \sin\frac{\theta}{2} \right),
\end{equation}
where $\theta$ is the angle between $\bi{r}$ and $\hat{\bi{x}}$.
Such an angular dependence indicates that the potential drops monotonously as $\theta$ increases from $0^\circ$ to $90^\circ$.
Therefore, the PM force bends the potential more along $\theta=90^\circ$ compared to $\theta=0^\circ$, allowing larger out-flowing probability current.
Another character of the angular dependence follows straightforwardly:
The ratio between the ionization rates along $\theta=90^\circ$ and $\theta=0^\circ$ decreases as $F_\text{PM}$ increases,
which is in line with the expectation that ionization in weaker fields is more sensitive to the barrier thickness.
\par

\subsection{Large-$\alpha_0$ and Large-$F_\mathrm{PM}$ Limits:\\ an Approximation using End-point Scaling}
\label{subsec:ScaledIonRate}

We note that for sufficiently large $\alpha_0$, the eigenstate of the field-free KH Hamiltonian exhibits the dichotomous structure:
the wave function concentrates near the two end points, forming two non-overlapping parts.
Hence, our interested region is in the vicinity of the end point $+\alpha_0 \hat{\bi{x}}$ or $-\alpha_0 \hat{\bi{x}}$,
where $r_+\ll\alpha_0$ or $r_-\ll\alpha_0$ is satisfied, and $V_0$ reduces to \cite{pont_radiative_1990}:
\begin{equation}
\begin{aligned}
    &V_0(\alpha_0; \bi{r}) \simeq \tilde{V}_0(\alpha_0;  \bi{r}_+)   &\text{for}\ r_+ \ll \alpha_0, \\
    &V_0(\alpha_0; \bi{r}) \simeq \tilde{V}_0(\alpha_0; -\bi{r}_-)   &\text{for}\ r_- \ll \alpha_0. \\
    \label{eq:EndPtVicinityApprox1}
\end{aligned}
\end{equation}
The ``end-point approximate'' potential $\tilde{V}_0$, is expressed as
\begin{equation}
    \tilde{V}_0(\alpha_0; \bi{r}) = -\frac{2}{\pi} \frac{1}{\sqrt{2\alpha_0 r}} \mathrm{K}\left(\sqrt{\frac{1-\hat{\bi{r}}\cdot \hat{\bi{x}}}{2}}\right).
    \label{eq:EndPtVicinityApprox2}
\end{equation}
\par

Under large-$\alpha_0$ limit, the ``end-point approximate'' dynamics can be further scaled to obtain $\alpha_0$-independent results.
Taking the end point $+\alpha_0 \hat{\bi{x}}$ as an example,
by introducing scaled coordinate variable $\bi{\rho}=\alpha_0^{-1/3}\bi{r}_+$,
the stationary Schrödinger equation [Eq.~(\ref{eq:TISE})] becomes
\begin{equation}
    \left[ -\frac12\nabla_\rho^2 + \tilde{V}_0'(\bi{\rho}) - \bi{F}'_\text{PM}\cdot\bi{\rho} \right]\varPhi = W\varPhi,
    \label{eq:ScaledTISE}
\end{equation}
where
\begin{equation}
    \tilde{V}_0'(\bi{\rho}) = - \frac{1}{\pi}\sqrt{\frac{2}{\rho}}\mathrm{K}\left( \sqrt{\frac{1-\hat{\bi{\rho}}\cdot \hat{\bi{x}}}{2}} \right)
\end{equation}
is the scaled approximate potential,
$W=\alpha_0^{2/3}E$ is the scaled energy,
$\bi{F}'_\text{PM}=\alpha_0 \bi{F}_\text{PM}$ is the scaled PM force,
and $\varPhi(\bi{\rho})=\alpha_0^{1/2}\varPsi(\bi{r}_+)$ is the scaled normalized wave function
(note that the origin has shifted to end point $+\alpha_0 \hat{\bi{x}}$).
Furthermore, we introduce the scaled time variable $\tau=\alpha_0^{-2/3}t$,
thus the TDSE Eq.~(\ref{eq:TDSE}) neglecting $\Lambda(t)$ is rewritten as
\begin{equation}
    \ii\pd_\tau\varPhi = \left[ -\frac12\nabla_\rho^2 + \tilde{V}_0'(\bm{\rho}) - \bi{F}'_\text{PM}\cdot\bi{\rho} \right]\varPhi.
    \label{eq:ScaledTDSE}
\end{equation}
Eqs.~(\ref{eq:ScaledTISE}) and (\ref{eq:ScaledTDSE}) are both $\alpha_0$-independent.
Finally, the ionization rate, which is of our interest, follows the scaling law
\begin{equation}
    \Gamma' = \alpha_0^{2/3}\Gamma.
\end{equation}
\par

\begin{figure*}
    \includegraphics[width=17.8cm]{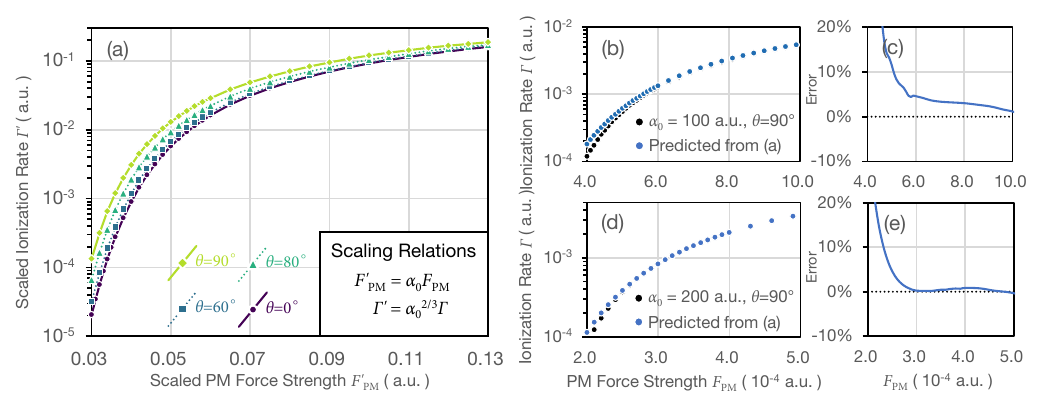}
    \caption{
        (a) Ionization rates obtained through end-point scaling.
        (b)(d) Comparison of actual ionization rates and those predicted from scaling,
        with corresponding errors shown in (c) and (e).
        }
    \label{fig:ScaledIonRate}
\end{figure*}

Fig.~\ref{fig:ScaledIonRate}(a) shows the scaled ionization rates obtained by solving the scaled TDSE
[Eq.~(\ref{eq:ScaledTDSE})] (smooth turn-on is also applied).
This variable scaling scheme provides a way of presenting $\alpha_0$-independent result,
which can be used to predict physical quantities of the actual system of arbitrary $\alpha_0$ as long as the end-point approximation is valid.
\par

To examine the validity of our scaling scheme,
we employ the data of the ionization rate of the scaled system to predict that of the real system.
The results for $\alpha_0=100\ \text{a.u.}$ and $\alpha_0=200\ \text{a.u.}$ at $\theta=90^\circ$ are shown in Figs.~\ref{fig:ScaledIonRate}(b) and \ref{fig:ScaledIonRate}(d),
with corresponding errors shown in Figs.~\ref{fig:ScaledIonRate}(c) and \ref{fig:ScaledIonRate}(e), respectively.
It is apparent that the predicted ionization rates match well the actual ones and match better for stronger PM forces.
What's more, at a certain $F'_\text{PM}=\alpha_0 F_\text{PM}$, larger $\alpha_0$ gives better match between predicted and actual ionization rates.
To explain this, we refer to the end-point approximation we made in the scaling scheme in Eq.~(\ref{eq:EndPtVicinityApprox1}),
where we assumed the condition $r_+\ll \alpha_0$ or $r_-\ll \alpha_0$.
We recast such a condition from two aspects.
For one thing, this condition requires the wave function to concentrate in the vicinity of two end points,
which is satisfied only for $\alpha_0>20\ \text{a.u.}$ (large-$\alpha_0$ condition).
For the other thing, this condition requires tunneling or over-barrier ionization to occur close to the end points,
where ``the location of ionization occurs'' can be characterized using the location of the tunneling exit or barrier top,
which is satisfied for large $F_\text{PM}$.
Therefore, the present scaling scheme is applicable in the large-$\alpha_0$ and large-$F_\text{PM}$ limit.
\par

\section{Impact of Ponderomotive Force on Multiphoton Ionization}
\label{sec:MPI}

Multiphoton ionization (MPI) plays a key role in adiabatic stabilization in intense and high-frequency laser fields
    \cite{gavrila_atoms_1992, pont_stabilization_1990, dorr_multiphoton_1991, pont_multiphoton_1991_i, pont_multiphoton_1991_ii}.
In this section we give a brief review on the multiphoton processes under the framework of HFFT and study the impact of PM force on it.
\par

Let us start from the space-translated Schrödinger equation in the KH frame,
where the Hamiltonian is given by Eq.~(\ref{eq:KH_Hamiltonian}).
Under the PM-free circumstance, the TDSE is written as
\begin{equation}
    \left[ \frac{\bi{p}^2}{2} + V\left(\bi{r}+\bi{\alpha}(\bi{r},t)\right) \right]\varPsi = \ii\pd_t\varPsi.
    \label{eq:PMFreeTDSE}
\end{equation}
The potential term $V(\bi{r}+\bi{\alpha}(\bi{r},t))$ is periodic, thus a Floquet-Fourier-form solution with quasi-energy $E$ can be sought \cite{gavrila_atoms_1992}:
\begin{equation}
    \varPsi(\bi{r},t) = \ee^{-\ii E t}\sum_{n=-\infty}^{+\infty} \varPsi_n(\bi{r}) \ee^{-\ii n \omega t}.
    \label{eq:FloquetFourierSolution}
\end{equation}
We also Fourier-analyze the potential, giving
\begin{equation}
    V(\bi{r}+\bi{\alpha}(\bi{r},t)) = \sum_{n=-\infty}^{+\infty} V_n(\alpha_0; \bi{r}) \ee^{-\ii n \omega t},
    \label{eq:FourierPotential}
\end{equation}
and the Fourier coefficients
\begin{equation}
    V_n(\alpha_0; \bi{r}) = \frac{1}{2\pi} \int_0^{2\pi/\omega} V(\bi{r}+\bi{\alpha}(\bi{r},t)) \ee^{\ii n \omega t} \dd t.
\end{equation}
Substituting Eqs.~(\ref{eq:FloquetFourierSolution}) and (\ref{eq:FourierPotential}) into Eq.~(\ref{eq:PMFreeTDSE})
gives an infinite system of coupled differential equations for $\varPsi_n(\bi{r})$.
The potential $V(\bi{r})$ has a Coulomb tail, which requires the solution to behave asymptotically as \cite{gavrila_atoms_1992}
\begin{equation}
\begin{aligned}
    \varPsi(\bi{r},t) \rightarrow & \frac{1}{(2\pi)^{3/2}} \ee^{-\ii E t}\\
                                    & \times  \sum_{n=-\infty}^{+\infty} \ee^{-\ii n \omega t} f_n(\alpha_0,\omega;\hat{\bi{r}}) \ee^{(\ii k_n r - \ln{(2 k_n r)}/k_n)}/r
\end{aligned}
\end{equation}
when $r\rightarrow\infty$, and the wave vectors follow the relation
\begin{equation}
    k_n =
    \begin{cases}
           \sqrt{2(E+n\omega)},     &\text{for}\ E+n\omega\geqslant 0,\\
        \ii\sqrt{2\abs{E+n\omega}}, &\text{for}\ E+n\omega < 0.\\
    \end{cases}
\end{equation}
Thus, for closed channels where $E+n\omega < 0$, $\varPsi_n(\bi{r})$ exponentially damps and vanishes at infinity;
while for open channels where $E+n\omega \geqslant 0$, $\varPsi_n(\bi{r})$ is an out-going spherical wave,
and $f_n(\alpha_0,\omega;\hat{\bi{r}})$ therefore represents the $n$-photon ionization amplitude with the corresponding
$n$-photon decay rate $\Gamma_n$ given by
\begin{equation}
    \Gamma_n = k_n \int \abs{f_n(\alpha_0,\omega;\hat{\bi{r}})}^2 \dd\hat{\bi{r}}.
\end{equation}
The total decay rate $\Gamma_\text{MPI}$ is subsequently the sum of $\Gamma_n$ over all open channels.
\par

An iteration scheme proposed by Gavrila and Kami\'{n}ski \cite{gavrila_free-free_1984} can be applied to study the solution of Eq.~(\ref{eq:PMFreeTDSE}).
The zeroth order iteration gives the familiar ``adiabatic stabilization'' with
\begin{equation}
    \varPsi_n(\bi{r}) = \varPsi_0(\bi{r})\delta_{0n},
\end{equation}
where $\varPsi_0(\bi{r})$ is the eigenstate of the dressed potential $V_0(\alpha_0;\bi{r})$,
as discussed above in Sec.~\ref{sec:PMI}.
The first-order iteration gives the expression of $f_n$:
\begin{equation}
    f_n(\alpha_0,\omega;\hat{\bi{r}}) = -\frac{1}{\sqrt{2\pi}} \Braket{\varPsi_{\bi{k}_n}^{(-)}|V_n|\varPsi_0},
    \label{eq:MPIAngDecayRate}
\end{equation}
where $\Ket{\varPsi_{\bi{k}_n}^{(\pm)}}$ denotes the scattering solution of
\begin{equation}
    \left[ \frac{\bi{p}^2}{2} + V_0(\alpha_0;\bi{r}) \right]\varPsi = (E_0+n\omega)\varPsi,
\end{equation}
and behaves as a plane wave with wave vector $\bi{k}_n$ (with the same direction as $\hat{\bi{r}}$) plus an outgoing ($+$) or incoming ($-$) spherical wave,
which is respectively normalized to the delta function in the momentum space.
It's also worth noting that in order for the iteration to converge and the first-order iteration to represent a good approximation,
the frequency of the laser needs to be sufficiently large, and more explicitly, $\omega\gg I_\text{p} = \abs{E_0}$ is needed,
where $E_0$ is the ground-state energy of potential $V_0$.
\par

In order to obtain the decay rate of MPI, we apply Eq.~(\ref{eq:MPIAngDecayRate}),
and replace $\Ket{\varPsi_{\bi{k}_n}^{(-)}}$ with a plane wave $\Ket{\bi{k}_n}$,
which is in fact the first order Born approximation.
The calculation is still carried out numerically on a Cartesian grid.
Under PM-free circumstances, $\varPsi_0$ is obtained by simply using imaginary-time evolution,
while for those with the PM force, as the eigenstate exhibits a continuum character beyond the downstream end point,
imaginary-time evolution is unable to give a ground state.
Under such circumstances, we use a trick of placing a wide absorbing boundary in the down-stream direction to suppress probability growth in the continuum region during the evolution.
By properly setting the parameter of the absorbing boundary, the imaginary-time evolution gives a satisfactory bound eigenstate, which exhibits a single-lobe structure.
\par

\begin{figure}
    \includegraphics[width=8.6cm]{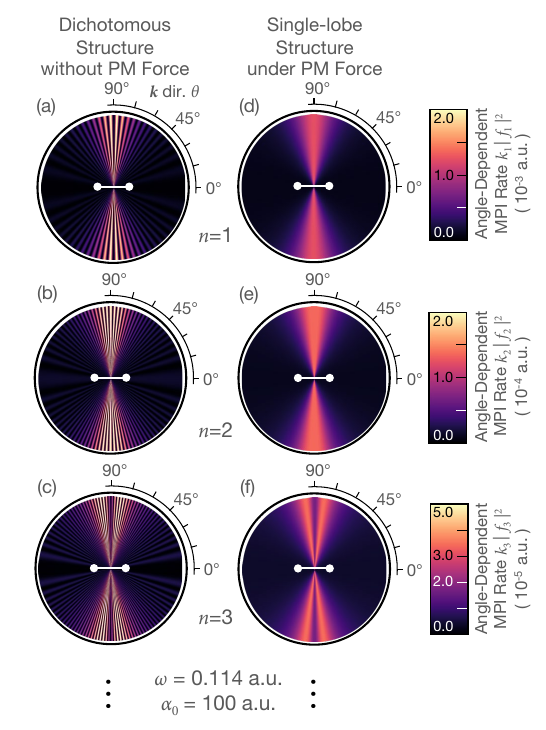}
    \caption{
        Angle-dependent multiphoton ionization rates $k_n \abs{f(\hat{\bi{r}})}^2$ from (a)-(c) PM-free and (d)-(f) PM-existing ground states,
        with $\omega=0.114\ \text{a.u.}$ and $\alpha_0=100\ \text{a.u.}$
        The first, second, and third rows display single-, double-, and triple-photon ionization rates, respectively.
    }
    \label{fig:MPIRate}
\end{figure}

Shown in Fig.~\ref{fig:MPIRate} are the angle-dependent MPI rates $k_n \abs{f(\hat{\bi{r}})}^2$ obtained with the above scheme.
The total ionization rate is little influenced by the presence of the PM force,
while the angular rate, which corresponds to the photoelectron momentum distribution,
has shown substantial structure alteration under the PM force.
Clearly, the PM-free case with an initial state that has a dichotomous structure exhibits interference fringes in the angular rates.
It was interpreted by Pont as resulting from the interference of electron waves ejected from two end points \cite{pont_multiphoton_1991_ii}.
However, the component of the PM force along the polarization direction results in a single-lobe structure in the initial state,
effectively closing one of the slits, thereby eliminating the fringes.
It is found that even a tiny PM force along the polarization direction is capable of breaking the dichotomy,
we can infer that for common laser frequency (visible light or UV) and high laser intensities ($\alpha_0 > 50\ \text{a.u.}$),
little portion of KH atoms can preserve their dichotomy geometry and contribute to the interference signal in a real laser field.
\par

\section{Ponderomotive-Force-Induced Tunneling/Over-Barrier Ionization Versus Multiphoton Ionization}
\label{sec:cp_PMI_MPI}

\begin{figure*}
    \includegraphics[width=17.8cm]{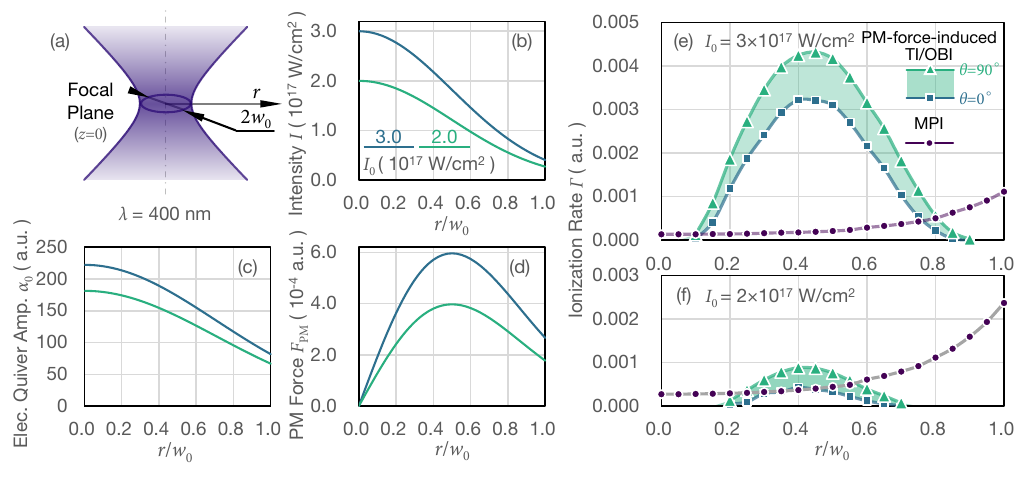}
    \caption{A comparison of PM-force-induced tunneling ionization (TI)/ over-barrier ionization (OBI)
             and multiphoton ionization (MPI) rates on the focal plane of a linearly polarized Gaussian beam.
             (a) The laser has a wavelength of $\lambda=400\ \text{nm}$,
             $w_0$ is the beam waist and $r$ denotes the radial distance from the focal axis.
             Two peak intensities $I_0=2\times 10^{17}$ W/cm$^2$ and $I_0=3\times 10^{17}$ W/cm$^2$ are used.
             (b)-(d) display the distribution of intensity, electron quiver amplitude, and the strength of the PM force on the focal plane, respectively.
             (e) and (f) show the comparison between PM-force-induced TI/OBI and MPI rates of lasers with two given peak laser intensities.
            }
    \label{fig:cp_IonRate}
\end{figure*}

In this section we take the Gaussian laser of Eq.~(\ref{eq:laser_profile}) as an example. We assign the laser parameter $\lambda=400\ \text{nm}$ ($\omega=0.114$ a.u.),
$w_0=17.5\ \mathrm{\mu m}$, with two peak intensities $I_0=2\times 10^{17}\ \text{W}/\text{cm}^2$ and $I_0=3\times 10^{17}\ \text{W}/\text{cm}^2$.
All the laser parameters as well as PM forces used in the calculations are accessible in experiments.
We note that such laser parameters have already been achieved \cite{wang_085pw_2017}.
For atoms in the focal plane ($z=0$) within one beam waist ($r \leqslant w_0$), the high-frequency criteria is well satisfied. Fig.~\ref{fig:cp_IonRate} displays the comparison of ionization rates of PM-force-induced ionization and MPI at different locations on the focal plane.
For a radial intensity distribution of Gaussian type, the PM force is pointing outwards along the radial direction,
with a strength increasing from zero to maxima as the radial location moves from zero to half beam waist ($r/w_0=0.5$),
and slowly decaying when moving further out [Fig.~\ref{fig:cp_IonRate}(d)].
Although the PM force peaks at half beam waist, ionization induced by the PM force peaks a bit inward, as clear from Figs.~\ref{fig:cp_IonRate}(e) and \ref{fig:cp_IonRate}(f).
The reason is that the laser intensity itself is greater near the beam axis, resulting in a larger $\alpha_0$ and lower $I_\text{p}$,
thereby effectively shifting the peak of PM-force-induced ionization to be inside the location of half beam waist.
Comparison of Figs.~\ref{fig:cp_IonRate}(e) and \ref{fig:cp_IonRate}(f) also reveals that as the intensity increases, MPI gets restrained due to the increase of $\alpha_0$,
while the PM-force-induced ionization grows exponentially and takes over as the dominant factor contributing to ionization, resulting in destabilization.
\par

\section{Conclusions}
\label{sec:concl}

Atomic stabilization in intense and high-frequency laser fields has attracted widespread attention and discussion for decades,
and is convinced to result in decreasing ionization probability with increasing laser intensity.
However, the present work indicates that as the laser intensity increases, the
PM force, which is induced by the non-uniform spatial distribution of laser intensity near the focal spot,
will not only break the symmetry of the well-known dichotomous structure,
but could also bring about tunneling and even over-barrier ionization,
resulting in breakdown of atomic stabilization.
Our work casts light on the importance of an improved model of atomic stabilization that accounts for the influence of the PM force.
\par

\begin{acknowledgments}
M. Z. would like to thank Zhaohan Zhang and Yaohai Song for helpful discussions and support during this research.
This work was supported by the National Natural Science Foundation of China (Grant Nos.~92150105, 11904103, 11974113) and the Science and Technology Commission of Shanghai Municipality (Grant Nos.~21ZR1420100, 19JC1412200).
Numerical computations were in part performed on the ECNU Multifunctional Platform for Innovation (001).
\end{acknowledgments}

\bibliography{ArticleBib.bib}

\end{document}